\documentclass{nature}
\usepackage{}
\usepackage{amssymb}
\usepackage{amsmath}
\usepackage{amsfonts}
\usepackage{graphicx}
\usepackage{caption}
\usepackage{subfig}
\usepackage{txfonts}
\usepackage{epsfig}
\begin{document}

\title{Nematic Metal and Antiferromagnetic Insulator on the Hexagonal Kagome Lattice }
\author{Heng-Fu Lin, Yao-Hua Chen, Hai-Di Liu, Hong-Shuai Tao, Wu-Ming Liu$^\star$}

\maketitle

\begin{affiliations}
\item
Beijing National Laboratory for Condensed Matter Physics,
Institute of Physics, Chinese Academy of Sciences,
Beijing 100190, China

$^\star$e-mail: wliu@iphy.ac.cn

\end{affiliations}

\begin{abstract}
Hexagonal Kagome lattice is a multiband system with a quadratic band crossing point, in contrast with honeycomb lattice with linear band crossing point, which has exotic correlated effect and can produce various novel quantum states. Here we investigate the phase diagram of the fermions on the hexagonal Kagome lattice as a function of  interaction, temperature and lattice anisotropy, by combining the cellular dynamical mean-field theory with the continuous time quantum Monte Carlo method. For weak interaction, the quadratic band-crossing point is broken to linear band crossing point and the system is the semi-metal. With the increasing of the interaction, the system goes a first order transition to antiferromagnetic Mott insulator at low temperature.  Below a critical temperature, due to the charge nematic fluctuation, a nematic metal forms between the weak coupling semi-metal and strong correlated Mott insulator. When the lattice anisotropy increases, the region of the nematic metal is enlarged. Furthermore, we discuss how to detect these phases in real experiments.
\end{abstract}

In recent years, many novel phases driven by the correlated effect have been found in two-dimensional multi-band Fermi systems with a band-crossing point, such as heavy-fermion behavior \cite{Kondo,Masafumi}, superconductivity \cite{Takada}, and spin liquid \cite{Lefbvre,Shimizu,Meng, Balents}. One of the studied case is the honeycomb lattice with linear band crossing point at the Fermi surface, in which the low energy physics can be described by a Dirac Fermion \cite{CastroNeto,Fadi,Xiaolong,chang,niu}. Another system, which has a quadratic band crossing point (QBCP), is also very important and interesting, such as  the hexagonal Kagome lattice \cite{Yao,Andreas}, and check board lattice \cite{sun1}. Many synthesis nano materials can be theoretically mapped to the hexagonal Kagome lattice, such as the triangular organic material $\kappa$-$BEDT(CN)_{3}$ \cite{Shimizu}, Kagome lattice herbertsmithite \cite{Matthew}, and three-dimensional hyperkagome lattice magnet $(Na_{4}Ir_{3}O_{8})$ \cite{Sasaki}. Various interesting novel phases are expected to found in this system, such as the chiral spin liquid \cite{Yao} with time-reversal symmetry spontaneously broken in the Kitaev model
, non-magnetic order \cite{Richter} in spin-half Heisenberg antiferromagnet, topological phase transition \cite{Andreas} and interaction-driven topological insulators \cite{Jun}.

The hexagonal Kagome lattice (or star lattice) is a ``cousin'' of both the honeycomb lattice \cite{Duan} and the Kagome lattice \cite{Ruostekoski}. It can be viewed as an ``interpolating'' lattice between
the honeycomb and the Kagome: if one shrinks the triangles at the
vertices of the underlying honeycomb lattice to their center
points, the honeycomb lattice is recovered, while expanding the
triangles until their corners touch produces the Kagome lattice. So
the frustration strength of this model is between the Kagome lattice
and the honeycomb lattice. The progress in experiments and theories of the optical lattice \cite{Jaksch,Hofstetter,Greiner,wu,zhou,liu,pu} might provide a promising way to simulate the model. Recently, a new antiferromagnetic material $[Fe_{3}(\mu_{3}$-$O)(\mu$-$OAc)_{6}$-$(H_{2}O)_{3}][Fe_{3}(\mu_{3}$-$O)(\mu$-$OAc)_{7.5}]_{2}¡¤7H_{2}O$ has been found to have a structure of hexagonal Kagome lattice \cite{Zheng}. Those make it possible to observe the many body phenomena existing in the systems with QBCP. The QBCP  is protected by time reversal symmetry and $C_{6}$ rotational symmetry in this model. However, it has  instabilities for interactions, leading to  quantum phase transitions.  Therefore, it is desirable to investigate the charge fluctuation with spins and the phase diagram on hexagonal Kagome lattice, which have not previously been studied.

In this work, we investigate the charge and magnetic order of the correlated fermion on the hexagonal Kagome lattice by employing the cellular dynamic-mean-filed theory \cite{Kotliar} combined with continue time quantum Monte Carlo method \cite{Rubtsov}. We obtained the phase diagrams  about the effect of interaction $U$, lattice anisotropy $\lambda$ and temperature $T$. The antiferromagnetic insulator (AFI) phase is found when $U>U_{c2}(T)$, which is indicated by an antiferromagnetic order and a finite charge gap.  A nematic metal (NM) phase forms between the AFI and the rotational symmetry broken semi-metal (SM) at $U<U_{c1}(T)$, which is induced by the charge nematic fluctuation \cite{Fradkin,Vojta}. This novel NM is identified by anisotropy momentum resolved spectra and non-Fermi liquid  behavior. The region of the NM phase can be enhanced by the lattice anisotropy. The critical temperature of  the phase transition to the NM is higher than the AFI. These interesting phases can be probed by the angle-resolved photoemission spectroscopy (ARPES) \cite{Damascelli}, neutron scattering \cite{Marshall}, nuclear magnetic resonance (NMR) \cite{Limot} and other experiments.

\section*{Results}

\subsection{The model and cellular dynamical mean field theory.}
We consider the hubbard model on hexagonal Kagome lattice with nearest-neighbor
hopping $t$ on vertex triangles and $t^{\prime}$ ($t^{\prime}$=$\lambda$$t$) between
triangles shown in Fig. 1(a). The Hamiltonian can be written as,
\begin{eqnarray}
H=-t\sum\limits_{\langle ij\rangle\sigma\Delta}c_{i\sigma}^{\dag}c_{j\sigma}-t^{\prime}\sum\limits_{\langle ij\rangle\sigma\Delta\rightarrow\Delta}c_{i\sigma }^{\dag}c_{j\sigma}
-\mu\sum\limits_{i\sigma}n_{i\sigma}+U\sum\limits_{i}n_{i\uparrow}n_{i\downarrow},
\end{eqnarray}
where $c^{+}_{i\sigma }$ and $c_{i\sigma }$ are the creation and the
annihilation operator of electrons with spin index $\sigma$ at site
$i$, $n_{i\sigma}=c_{i\sigma }^{+}c_{i\sigma }$ is the density
operator. $U$ is the on-site repulsion interaction, and $\mu$ is the chemical
potential. The Hubbard model respects $SU(2)$ spin symmetry, time reversal symmetry and $C_{6}$ lattice rotational symmetry.

In the hexagonal Kagome lattice, there are six sites in each unit cell with
two sublattice $A$ and $B$ (see Fig. 1(a)). The super-lattice vectors in the
real-space (see the green arrows in Fig. 1(a)) and vectors in recipe space (see the red arrows in Fig. 1(b)) are shown. The reduced first Brilioun zone is shown in Fig. 1(b), which
is identical to honeycomb lattice and Kagome lattices. Firstly, we consider the noninteracting case $U/t=0$. By using an
orthogonal transformation, the kinetic term of the Hamiltonian is
diagonalized at each $k$, and we get the noninteracting energy
dispersion with six mini bands. The energy spectrum is $\varepsilon_{1,2,4,5}(k)=-\frac{t}{2}\pm (\frac{9}{4}t^{2}+t^{\prime2}\pm tt^{\prime}\sqrt{3+2cosk_{1}+2cosk_{2}+2cos(k_{1}-k_{2})})^{1/2}$ and
$\varepsilon_{3,6}(k)=t\pm t^{\prime}$, where $k_{1}=(2+\sqrt3)k_{x}$, $k_{2}=(1+\frac{1}{2}\sqrt3)k_{x}+(\frac{3}{2}+\sqrt3)k_{y}$.

For $t^{\prime}<\frac{3}{2}t$ case, the  bands
$\varepsilon_{3}(k)$ and $\varepsilon_{6}(k)$ form flat band over
the whole Brillouin zone. The dispersive bands $\varepsilon_{1}(k)$ and
$\varepsilon_{2}(k)$ ($\varepsilon_{4}(k)$ and $\varepsilon_{5}(k)$) touch each other at $K$ and $K^{\prime}$  with
linear dispersion,  which  named  the Dirac point.
However, the band $\varepsilon_{3}(k)$ and
$\varepsilon_{4}(k)$ ($\varepsilon_{6}(k)$ and
$\varepsilon_{5}(k)$ ) touch each other at $\Gamma$ with quadratic
dispersion, which named the quadratic band crossing point (QBCP).  For
$t^{\prime}>\frac{3}{2}t$,  there is a little different. The
flat band $\varepsilon_{4}(k)$ and the dispersive band
$\varepsilon_{5}(k)$ touch each other at $\Gamma$ with quadratic
dispersion, and  form the quadratic band crossing point (QBCP).
To see the quality more clearly, we calculate the band structure
along the high symmetry point along the $ M $ $\Gamma $ $ K $ and $
M $ direction (Fig. 2(b)). There are Dirac points at $K$ and $K^{\prime}$,
quadratic band crossing point (QBCP) at $\Gamma$ point.

We can define the density of states (DOS) including the chemical
potential, as $\rho(\omega)=(1/N)\Sigma_{k,\alphaup}1/(\omega+\mu-\varepsilon_{\alphaup}(k))$.
And here we define $G=\sum_{k}\frac{1}{w+\muup-t(k)}$, using the hopping matrix, we calculate the non-interacting
density of state,according to the formula
$\rho_{0}(\omegaup)=-\frac{1}{\pi}ImG_{ii}(w+i0^{+})$. At half
filling, there is a $\delta $-function peak appear at the Fermi
surface due to the flat band, four van-hove singulate and a huge
energy gap just below the Fermi surface (Fig. 1(d)).

In this paper, we study the QBCP's quantities in this system, and set the system filling factor $1/2$ and hopping amplitude $t^{\prime}<\frac{3}{2}t$ firstly. In order to include the short range interaction, we use the cellular dynamical mean field theory, which is the cluster extention of DMFT.
The dynamical mean-field theory (DMFT) has given substantial theoretical progress in understanding the Mott transition, but this method, not treating spatial fluctuation. Within the cellular dynamical mean field theory, we can tract the frustration and nonlocal interaction more efficiently. The Mott transition and magnetic properties on the full frustrated Hubbard model with different
lattices structure, the relax of the frustration
influence on the mott transition \cite{Yuta}, heavy-fermion
behavior on the frustrated Kagome Hubbard model\cite{Masafumi}.

\subsection{Finite temperature phase diagram with the isotropic hopping.}
The finite temperature phase diagram of half-filling isotropic ($t^{\prime}=t$) Hubbard model is shown in Fig. 2. At low temperature, such as $T<T_{1c}\sim 0.1$, three phases are formed at the different interaction strength $U$. When $U<U_{c1}\sim 3.4$ at $T=0.05$, the system is semi-metal (SM) with two Dirac points broken from the QBCP, which can be described by the Dirac fermion very well. When $U>U_{c2}\sim 5.2$ and $T=0.05$ the system become an insulator phase with antiferromagnetic order through a first-order phase transition. This phase is called antiferromagnetic insulator (AFI). A nematic metal (NM) \cite{Lawler} emerges at intermediated interaction $U_{c1}<U<U_{c2}$, where $U_{c1}$ and $U_{c2}$ are critical points. In this region, the system is metallic with finite density of states in the Fermi level and anisotropic momentum-resolved single particle spectral.  With the increasing of the temperature, the nematic order in the metal side and antiferromagnetic order in the insulator side will be destroyed by the thermal fluctuation gradually.   At high temperature $T>T_{c2}$, the nematic order and antiferromagnetic order have been both broken by the thermal fluctuation, the system goes from semi-metal (SM) crossover to paramagnetic insulator (PMI).  The transition temperature of the nematic oder (green line with circles) and antiferromagnetic order (black line with squares) as a function of the interaction are carefully calculated.  The nematic and antiferromagnetic phase are not broken simultaneously. The antiferromagnetic order begin to be destroyed  at a start point with  critical temperature at $T_{c2}\approx 0.1$ and interaction $U_{c2}\approx 6.0$. The nematic order are completely destroyed at an end point with critical temperature $T_{c1}\approx0.135>T_{c2}$ and critical interaction $U_{c1}\approx6.2>U_{c2}$.

\subsection{Non-fermi liquid behavior and the nematic metal.}
The imaginary part of the on-site cluster self-energy $Im\Sigma_{11}(i\omega_{n})$ and propagator $ImG_{11}(i\omega_{n})$ provides information about the possible Fermi-liquid or non-Fermi liquid behavior of the system as well as the nature of the charge gap opening. In Fig. 3, we present $Im\Sigma_{11}(i\omega_{n})$ and $ImG_{11}(i\omega_{n})$ as a function of the Matsubara frequencies for different values of $U$ at $T=0.05$ and $\lambda=1.0$. For $U<U_{c1}\sim3.4$ and small $\omega_{n}$, $Im\Sigma_{11}(i\omega)\sim i\omega_{n}$, which is similar to the Fermi liquid.
For $U>U_{c2}\sim5.2$ the behavior is clearly different, the $Im\Sigma_{11}(i\omega_{n})$ and $ImG_{11}(i\omega_{n})$ increase when $\omega \rightarrow 0$, which implies a gap existing near the Fermi energy in the density of states.
For $U_{c1}<U<U_{c2}$, the $Im\Sigma_{11}(i\omega_{n})$ indicates a finite limiting value when $\omega_{n} \rightarrow 0$, which implies a finite lifetime for states near the Fermi energy, and $ImG_{11}(i\omega_{n})$ decrease to a small constant. In this regime, the behavior of $Im\Sigma_{11}(i\omega_{n})$ and $ImG_{11}(i\omega_{n})$ display exotic properties, which means the system translates to a novel metallic phase called nematic metal phase. The nematic state can be obtained from a Pomeranchuk \cite{Pomeranchuk} instability generated by forward scattering interactions in a normal metal, in which time reversal symmetry and spacial rotational invariance $C_{6}$ is broken spontaneously.

In the nematic phase region, the broadening of  the single particle spectral varies in momentum space. To illustrate this, the momentum-resolved spectral weight at the fermi level $A(k,\omega=0)$ in the first Brillouin zone is investigated for different values of interaction $U$ and temperature $T$. $A(k,\omega=0)$ is obtained by $A(k,\omega=0)\approx -\frac{1}{6\pi} \sum_{i=1}^{6}\lim_{\omega_{n }\to 0}ImG_{ii}(k,i\omega_{n})$, where $i$ is the site index within the chosen cluster. When $U=0$, the Fermi surface $A(k,0)$ is a point at $\Gamma$, where the flat band $\varepsilon_{3}(k)$ and the dispersive bands
$\varepsilon_{4}(k)$ touches each other with quadratic dispersions. When the interaction $U$ is weak, such as $U<U_{c1}$, the Fermi surface $A(k,0)$ breaks to two points, which implies that the system is a semi-metal. Fig. 4(a1) shows the Fermi surface $A(k,0)$ at $U=1.0, T=0.067$. The instable QBCP in 2D split into two Dirac points. This means the system break the $C_{6}$ rotational symmetries to $C_{2}$ spontaneously, which is also found in Ref. \cite{Sun2,Terletska,Park}. When the interaction $U>U_{c1}$, the broadening of the peak in $A(k,0)$  are away from the Brillouin-zone diagonal, the Fermi surface $A(k,0)$ becomes anisotropic (see Fig. 4(b2)) at $U=4.0, T=0.067$. The anisotropic Fermi surface, which have been studied by means of CDMFT approach \cite{Yuta,Okamoto}, indicates that the system become a nematic metal.

\subsection{Mott transition and antiferromagnetic insulator.}
In order to investigate the evolution of single particle spectral in the phase transition more clearly, we calculate the density of states (DOS) for different $U$ when $T=0.067$. The DOS is defined as $\rho(\omega)=-\frac{1}{6\pi}\sum_{l=1}^{6}ImG_{ll}(\omega+i\delta)$, where $l$ is the site index within cluster. The DOS can be derived from the imaginary time Green's function $G(\tau)$ by using maximum entropy method \cite{Jarrell}.
When $U=1.0$, in the semi-metal region, the DOS is zero in the Fermi level (see Fig. 4(b1)), and a sharp spectral weight peak is formed at the low energy region. When $U/t=4.0$, corresponding to the nematic metal, a pseudo-gap forms near the Fermi level (see Fig. 4(b2)). The sharp spectral weight peak  shifts to a higher energy gradually when the interaction increases. When $U=6.0$, in the gapped insulator region, the sharp spectral weight peak is suppressed by interaction and a gap is opened near Fermi level ( Fig. 4(b3)).

The double occupancy $D_{occ}=\partial E_{g}/ \partial U$ directly describes a quantum phase
transition tuned by $U$, where $E_{g}=\langle H \rangle/N$ is the ground state energy per site. The $D_{occ}$ as a function of $U$ is obtained by
$D_{occ}= \frac{1}{N_{c}}\sum_{i}^{N_{c}}\langle n_{i\uparrow}n_{i\downarrow}\rangle$ (see Fig. 5). We find that $D_{occ}$ decrease smoothly at the critical point $U_{SM-NM}$. However, a discontinuity can be observed near the critical point $U_{NM-AFI}$, which signals a first-order phase transition.

In order to investigate the evolution of magnetic order on the hexagonal Kagome lattice, we define the antiferromagnetic order parameter
$m=\frac{1}{N_{c}}\sum_{i}sign(i)(n_{i\uparrow}-n_{i\downarrow})$, where $i$ is the site index (see Fig. 1 (a)).
$sign(i)$ is defined as $sign(i)=1$ for site $a_{1}$, $a_{3}(b_{3})$, $b_{2}$ and $sign(i)=-1$ for site $b_{1}$, $b_{3}(a_{3})$, $a_{2}$.
Fig. 6 shows the evolution of the antiferromagnetic order parameter $m$ and charge gap $\Delta E$ as a function of interaction strength $U$, at $\lambda=1.0$ and temperature $T=0.05$. We find an antiferromagnetic insulator phase when $U>5.2$, in which the magnetic order appears simultaneously with the charge gap. This means that the magnetic order enhance the localization of electron on the lattice sites.  Two equivalent spin configuration of antiferromagnetic insulator can be found in insert (b) and (c) in Fig. 6. It is different from the previous studies in the unfrustrated system with flat band and a QBCP, which favor a Nagaoka ferromagnetic phase \cite{Park1} and nematic ferromagnetic. The insert (a) of Fig. 6 shows that there is no magnetic order forms when $T=0.2$.

\subsection{The phase diagram with the anisotropic hopping.}
We now turn to investigate the effect of anisotropic $\lambda=t^{\prime}/t$ on hexagonal Kagome lattice. The anisotropic $\lambda$ lead to the effective hopping amplitude intra trangle lattice and inter trangle lattice very different.  For $\lambda<1$, the electrons are more iternerant within every trangle; for $\lambda>1$, the electrons are more iternerant in intra trangle. When we consider the interaction, the electrons iternerant are great changed, the nematic metal phase are enhanced. We carefully calculate the momentum-resolved spectral, onsite self-energy and onsite green funciton. At the low
temperature regime,the magnetism and mott phase transition are greatly affected.

Fig. 7 shows the phase diagram about anisotropy and interaction when $T=0.067$. The phase boundary is shifted by the anisotropy $\lambda$. We can find that $U_{c1}=5.0$, $U_{c2}=6.2$ when $\lambda=1.2$ and $U_{c1}=4.2$, $U_{c2}=5.0$ when $\lambda=0.8$. The region of nematic metal phase is enlarged when $\lambda$ varies. This means the nematic metal become stable due to the enhancing of anisotropy. This provide a way to realize such state in the materials  and  to detect this phase in real experiments, such as under high pressure or distortion.

\section*{Discussion}

Besides being of fundamental theoretical interest, these phases of this model is of
broad experimental relevance. The nematic metal, which is similar as the  electronic nematicity
been observed in the iron pnictide \cite{Zhao,Kasahara} and copper oxide \cite{Hinkov,Daou} high temperature
superconductors, can be found in  other strong correlated  materials and can be probed by the angle-resolved photoemission spectroscopy (ARPES) \cite{Damascelli}, neutron scattering, nuclear magnetic resonance (NMR) \cite{Limot}. The antiferromagnetic insulator phase  been found in  the polymeric iron(III) acetate $[Fe_{3}(\mu_{3}$-$O)(\mu$-$OAc)_{6}$-$(H_{2}O)_{3}][Fe_{3}(\mu_{3}$-$O)(\mu$-$OAc)_{7.5}]_{2}¡¤7H_{2}O$ \cite{Zheng} of the same lattice structure can be detected  using neutron scattering, nuclear magnetic resonance (NMR) \cite{Limot}. In ultracold atom systems, both interaction and the hopping amplitude can be fully contralled using Feshbach resonenced and changing the lattice depth  respectly. Therefore, this model can be realized and these phases can also be detected using cold atom detecting technology, such as time of flight.

In summary, we have derived the phase diagram of correlated fermions on the hexagonal Kagome lattice as a function of temperature, interaction and anisotropy.  The nematic metal with non-Fermi liquid behavior is found when $U_{c1}<U<U_{c2}$ (such as $U_{c1}=4.3$, $U_{c2}=5.2$ when $T=0.067$), above a semi-metal phase at $U<U_{c1}$. The nematic metal breaks the time symmetry and $C_{6}$ symmetry and the phase region enlarged when the lattice anisotropy increases. When $U>U_{c2}$ (such as $U_{c2}=5.6$ when $T=0.067$), the system undergoes a first order phase transition from the gapless nematic metal to gapped antiferromagnetic insulator. The nematic metal and antiferromagnetic insulator phase are broken by thermodynamic fluctuation when $T>0.135$. Our studies provide a helpful step for understanding the nematic fluctuation and magnetic fluctuation in metal-insulator transition, which are  of relevance to the magnetic order phase and  charge nematic phase in the multiband systems with a QBCP.

\section*{Methods}
The Cellular dynamical mean field theory (CDMFT) is an extension of dynamical mean field theory (DMFT), which is able to partially cure DMFT's spatial limitations. We replace the site-impurity by a cluster of
impurities embedded in a self-consistent bath. Short-ranged spatial correlation are in this way treated exactly
inside the cluster, and a first momentum-dependence of the properties of the system is recovered. The cluster-impurity
problem embedded in a bath of free fermions can be written in a quadtratic form£¬
\begin{equation}
S_{eff}=\int^{\infty}_{0}\sum_{ij\sigma}c_{i\sigma}^{\dag}\mathcal{G}^{-1}_{ij\sigma}(\tau)c_{j\sigma}+U\int_{0}^{\beta}dt\sum_{i}n_{i\uparrow}(\tau)n_{i\uparrow}(\tau),
\end{equation}
and the $\mathcal{G}^{-1}_{ij\sigma}$ is the Weiss field.

Within CDMFT, the interacting lattice Green's function in the
cluster site basis is given by,
\begin{equation}
G^{-1}_{ij\sigma}(i\omega_{n})=\sum_{k}[i\omega_{n}+\muup-t(k)-\Sigma_{\sigma}(i\omega_{n})]^{-1}_{ij},
\end{equation}
where $\omega_{n}=(2n+1)\pi T$ are Matsubara frequenies, $\muup$ is
the chemical potential and $\hat{t}(\mbox{\boldmath $k$})$
is the Fourier-transformed hopping matrix for the super lattice. Here, we choose  the
unit cell as the cluster and  divide the two-dimensional lattice
into clusters consisting of six sites (Fig. 1(a)). And
the hopping matrix $t(k)$ of the cluster can be written as follows
($k\in RBZ$),
\begin{equation}
 \mathbf{}
 \left[\begin{array}{cccccc}
 0 & t & t & 0 & t^{\prime}*e^{-ik\cdot \delta(3)} & 0 \\
 t & 0 & t & 0 & 0 & t^{\prime}*e^{-ik\cdot \delta(4)} \\
 t & t & 0 & t^{\prime} & 0 & 0 \\
 0 & 0 & t^{\prime} & 0 & t & t \\
 t^{\prime}*e^{-ik\cdot \delta(1)} & 0 & 0 & t & 0 & t \\
 0 & t^{\prime}*e^{-ik\cdot \delta(2)} & 0 & t & t & 0
 \end{array}\right],
 \end{equation}
where $\delta(1)=\vec{m}=( 1+\frac{\sqrt3}{2}, \frac{3}{2}+\sqrt3),
\delta(2)=\vec{m}-\vec{n}=(-1-\frac{\sqrt3}{2}, \frac{3}{2}+\sqrt3),
\delta(3)=-\vec{m}=(-1-\frac{\sqrt3}{2},-\frac{3}{2}-\sqrt3) ,
\delta(4)=\vec{m}-\vec{n}=( 1+\frac{\sqrt3}{2},-\frac{3}{2}-\sqrt3)$ are
the nearest-neighbor super-lattice vectors. Given the Green function
for the effective medium, $\mathcal{G}^{-1}_{\sigma}$, we compute the cluster
Green function $G_{\sigma}$ and the cluster self-energy
$\Sigma_{\sigma}$, Where $\mathcal{G}^{-1}_{\sigma}$, $G_{\sigma}$ and
$\Sigma_{\sigma}$ are described by $6\times6$ matrices. The effective medium
$\mathcal{G}^{-1}_{\sigma}$ is then computed via the Dyson equation,
\begin{equation}
\mathcal{G}^{-1}_{\sigma}(\omega)=[\sum_{k}\frac{1}{\omega+
\muup-t(k)-\Sigma_{\sigma}(\omega)}]^{-1}+\Sigma_{\sigma}(\omega).
\end{equation}
We iterate the DMFT self-consistent loop until the convergence of
this procedure is achieved within $40$ iterations at most.

In each iteration, in order to solve the effective cluster model and to calculate $G_{\sigma}$ and $\Sigma_{\sigma}$, we
use the weak coupling interaction expansion continue time quantum mente carlo method (CTQMC)
\cite{Rubtsov}. At low $T$, the CTQMC method has the sign problem
for this system even in the half filling, due to the absence of the
particle-hole symmetry. We typically use $1.92\times10^{7}$ QMC sweeps to reach
sufficient computational accuracy at low temperature.


\begin{addendum}

\item [Acknowledgement]
H. F. Lin acknowledges very helpful discussions with N. H. Tong, D. X. Yao and W. Wu .
This work was supported by the NKBRSFC under grants Nos. 2011CB921502, 2012CB821305, NSFC under grants Nos. 61227902 and 61378017.

\item [Author Contributions]
H. F. L. performed calculations.
F. F. L., Y. H. C., H. D. L, H. S. T., W. M. L. analyzed numerical results.
H. F. L., Y. H. C., W. M. L. contributed in completing the paper.

\item [Competing Interests]
The authors declare that they have no competing financial interests.

\item [Correspondence]
Correspondence and requests for materials should be addressed to Heng-Fu Lin and Wu-Ming Liu.

\end{addendum}

\clearpage

\newpage
\bigskip

\textbf{Figure 1 The structure of hexagonal Kagome
lattice and its qualities in the non-interacting limit.}
(a) The structure of the hexagonal Kagome
lattice. The solid circles denote the A-sites and the open circles
denote the B-sites. The unit cell with the sites $a_{1}$,  $a_{2}$, $a_{3}$, $b_{1}$, $b_{2}$ and $b_{3}$  is indicated by the dashed rectangle shape. The purple arrows show real lattice vectors $m$ and $n$. (b) The reciprocal space vectors $k_{1}$, $k_{2}$ and first Brillouin zone (shown by yellow color) of the hexagonal Kagome lattice. (c) The band structure along the path is shown in Fig. 1(b). There are Dirac points at $K$ and $K^{\prime}$(not shown)
and QBCP at $\Gamma$. When the filling $f=1/2$, the phase
diagram involving QBCP only. (d) The non-interacting density of states
(DOS) of the system at half filling
$f=1/2$. There are four van-hove singular and two $\delta$
peak due to the flat band.

\textbf{Figure 2 Finite temperature diagram with the isotropic  hopping ($\lambda=1.0$).}
Phase diagram for isotropic  ($\lambda=1.0$)
Hubbard model on the hexagonal Kagome lattice as a function of $U$
and $T$. There are four phases in the phase diagram: (i) semi-metal (SM), (ii) antiferromagnetic insulator (AFI), (iii) nemetic metal (NM), (iv) paramagnetic insulator (PMI). The critical interaction and  critical temperature of the SM-NM phase transition are shown as ($U_{c1},T_{c1}$) . The critical points of  the phase transition to AFI are shown as ($U_{c2},T_{c2}$.

\bigskip

\textbf{Figure 3 Evolution of on-site self-energies and propagator.}
 Enhancement of on-site self-energies across the semi-metal and nematic metal transition with a increasing of $U$. The imaginary part of the on-site propagator $G_{11}(i\omega_{n})$ and on-site self-energy $\Sigma_{11}(i\omega)$ (inset) are plotted for different values of $U$ when $T=0.05$.

\bigskip

\textbf{Figure 4 Moenmentum-resolved spectral
weight $A(k,0)$ and local density of state.}
Left panel: Moenmentum-resolved spectral
weight $A(k,0)$ at the Fermi level in the first Brillouin zone when $T=0.067$. (a1) The semi-metal phase when $U/t=1.0$. (a2) The nematic metal phase when $U/t=1.0$. Right panel: The density of states (DOS) $\rho(\omega)$
at various interaction when temperature $T=0.067$. (b1) The zero density in the Fermi level shows the system is a semi-metal when $U/t=1.0$. (b2) When $U/t=4.0$, the system stays at the nematic metal phase, in which there is a finite density in the Fermi level. (b3) An obvious gap is opened when $U/t=6.0$, which means the system is an insulator.

\bigskip

\textbf{Figure 5 The evolution of double occupancy $D_{occ}$.}
The evolution of double occupancy $D_{occ}$ as a
  function of  $U$ for various $T$. The solid arrows show the critical points of semi-metal and nematic metal. And the critical points of nematic metal and antiferromagnetic insulator are shown by dash arrows.

\bigskip

\textbf{Figure 6 The evolution of the antiferromagnetic order parameter $m$ and the single particle gap $\Delta{E}$.}
The evolution of the antiferromagnetic order parameter $m$ and the single particle gap $\Delta{E}$ as a function of $U$ when $\lambda=1.0$ and $T=0.05$. Inset: (a) The evolution of $m$ and  $\Delta{E}$ as a function of $U$ at $\lambda=1.0$ and $T=0.2$. (b), (c) Two equivalent spin configuration of the antiferromagnetic insulator.

\bigskip

\textbf{Figure 7 The phase diagram with the anisotropic hopping  $\lambda\neq 1$.}
The phase diagram of
Hubbard model on hexagonal Kagome lattice as a function of $U$ and lattice anisotropic $\lambda$ when $T=0.067$. The blue line shows translation line of semi-metal and nematic metal, in which the circles indicate critical points. The translation line of the nematic metal and antiferromagnetic insulator is shown by black line, in which the critical points are signed by the square.

\newpage
\begin{figure}
\begin{center}
\epsfig{file=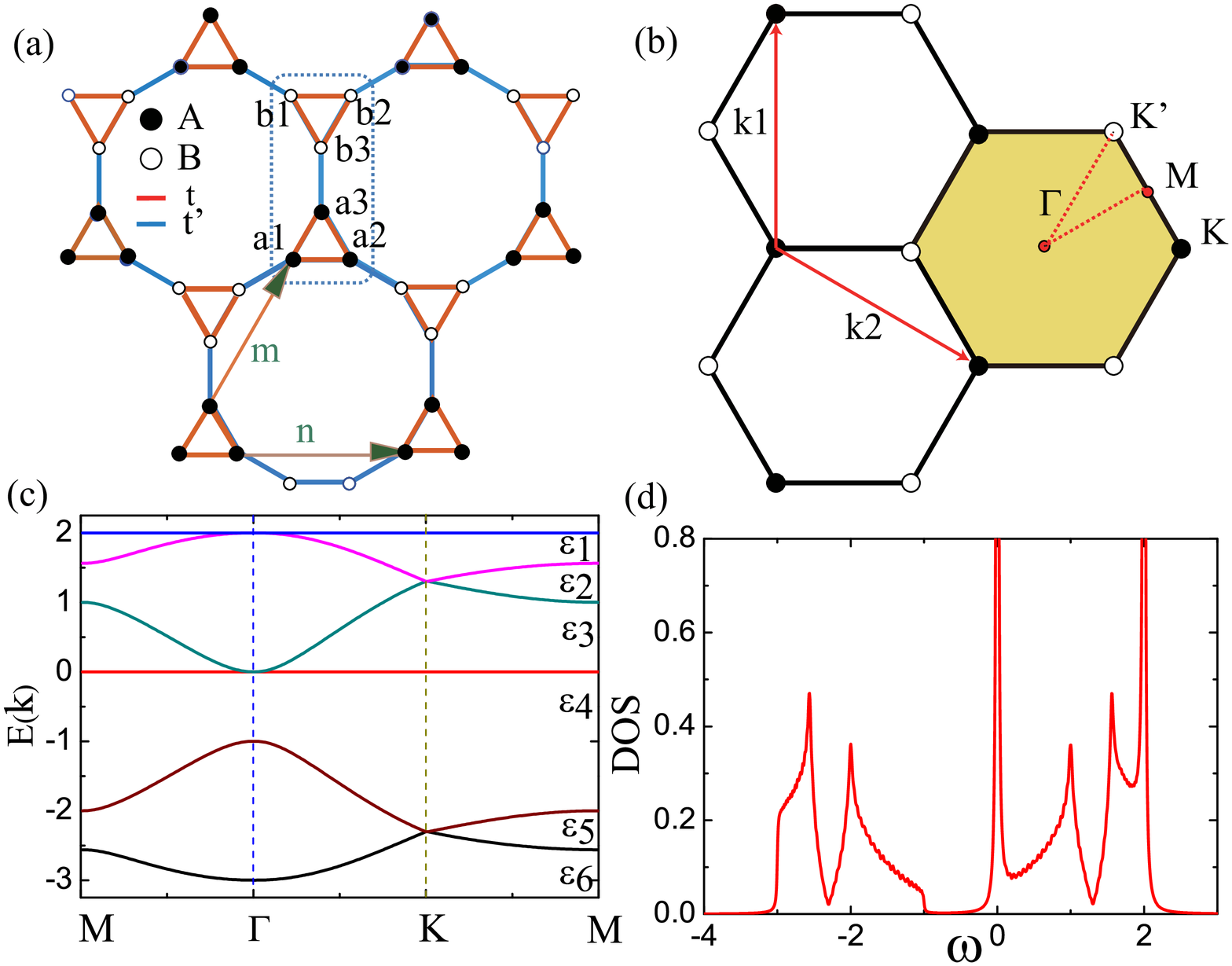,width=15cm}
\end{center}
\label{fig:TQPT}
\end{figure}
\begin{figure}
    \begin{center}
        \epsfig{file=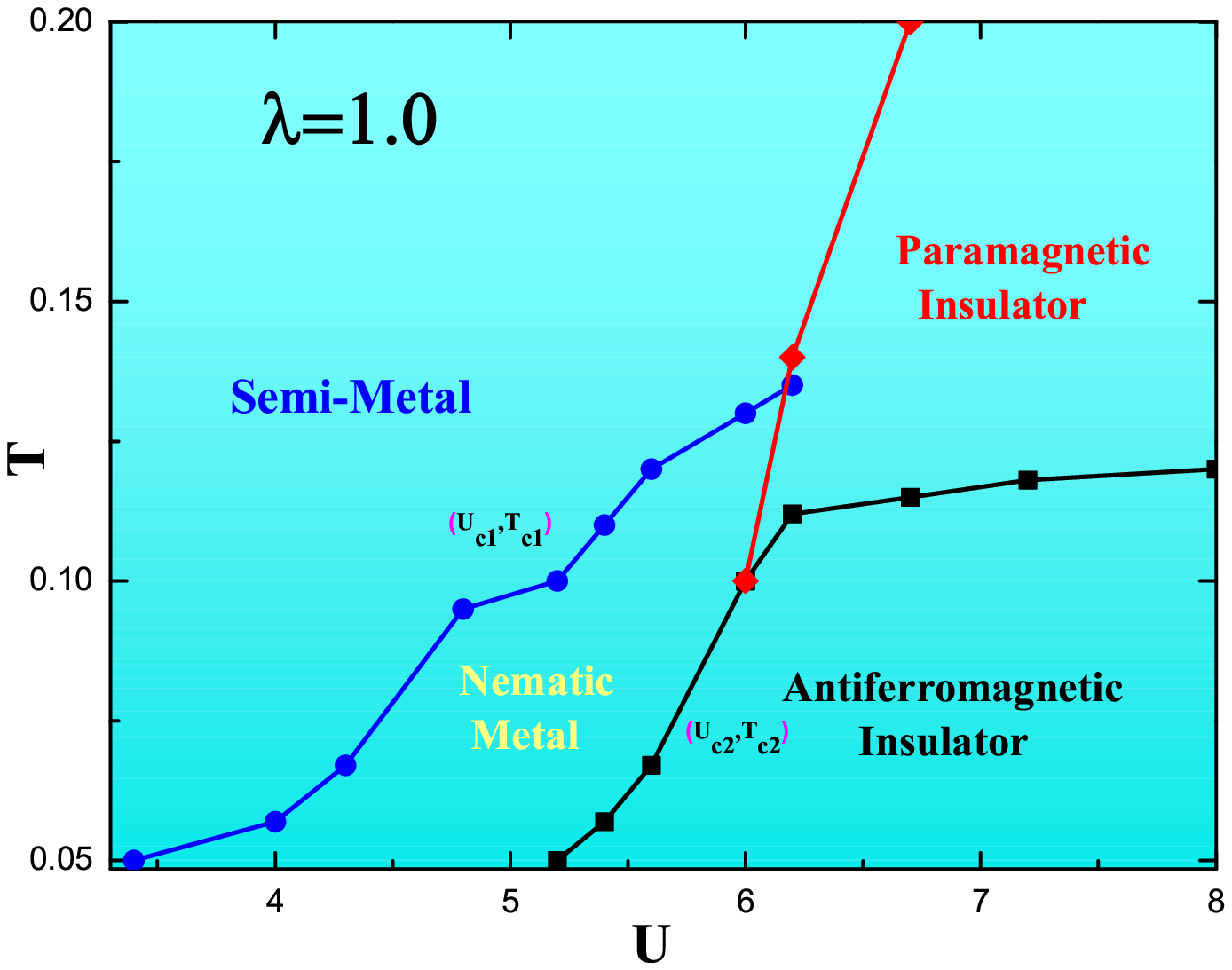,width=15cm}
    \end{center}
    \label{fig:Geo}
\end{figure}

\begin{figure}
    \begin{center}
        \epsfig{file=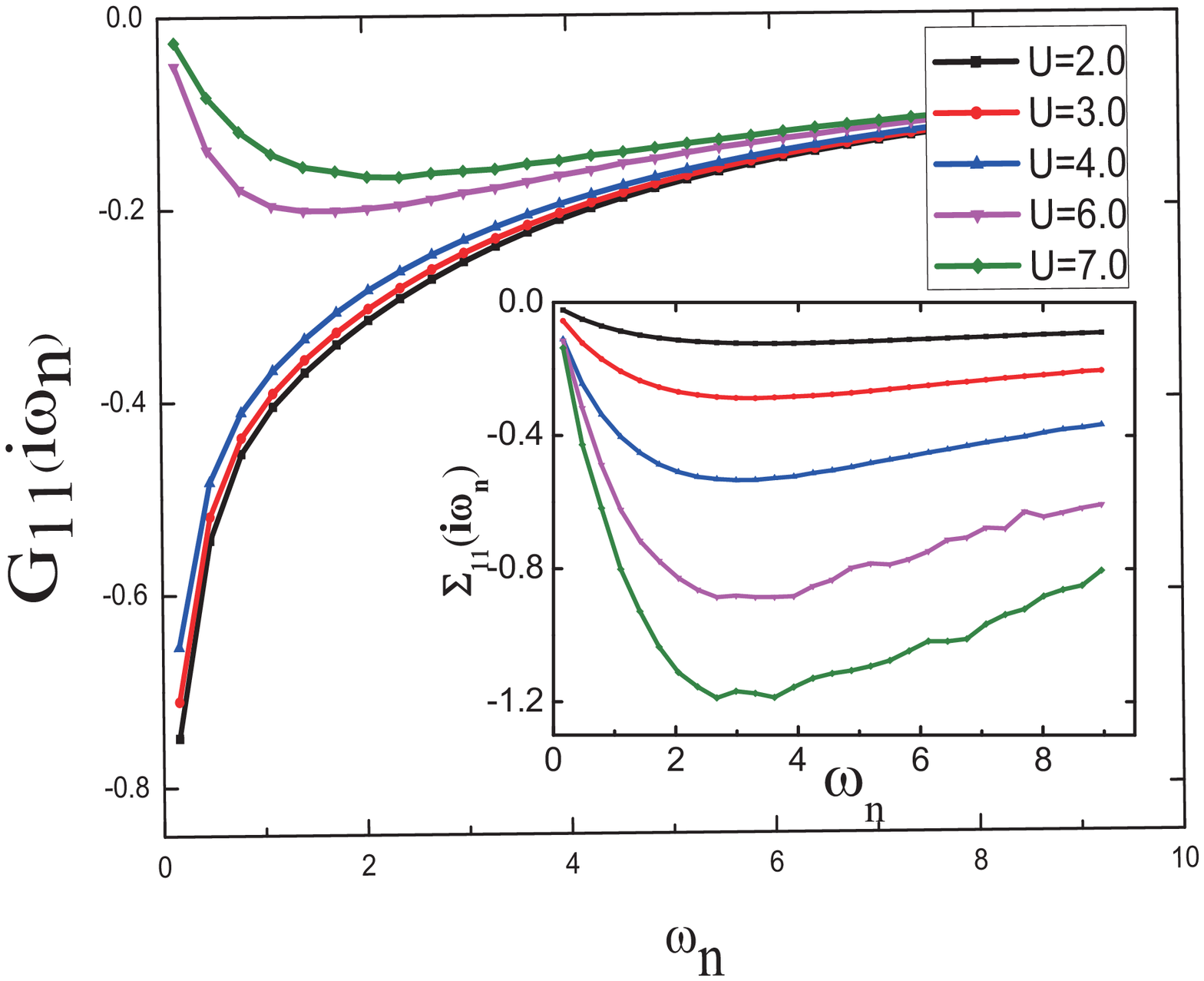,width=15cm}
    \end{center}
    \label{fig:DOS}
\end{figure}

\begin{figure}
    \begin{center}
        \epsfig{file=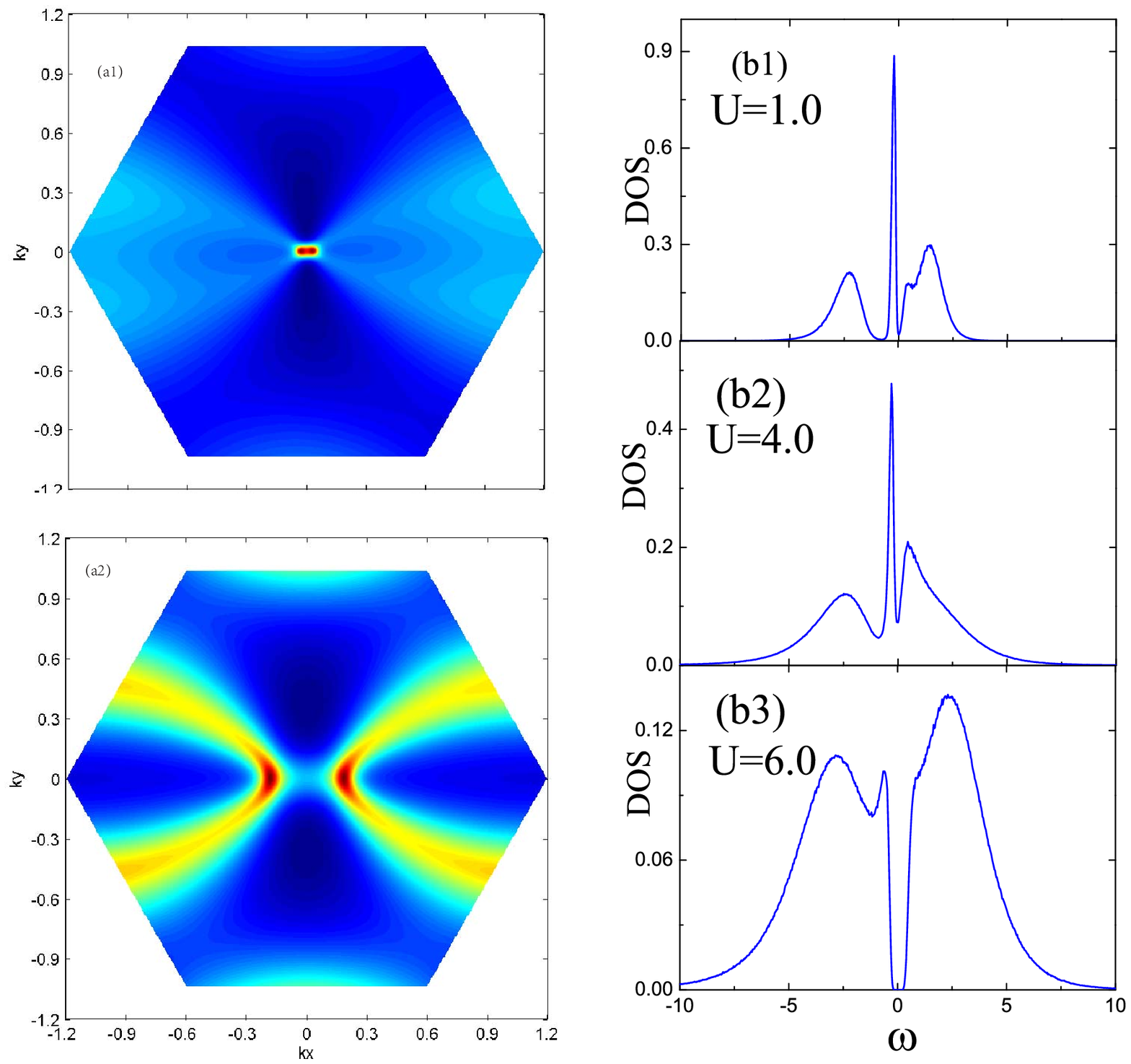,width=12cm}
    \end{center}
    \label{fig:TB}
\end{figure}

\begin{figure}
    \begin{center}
        \epsfig{file=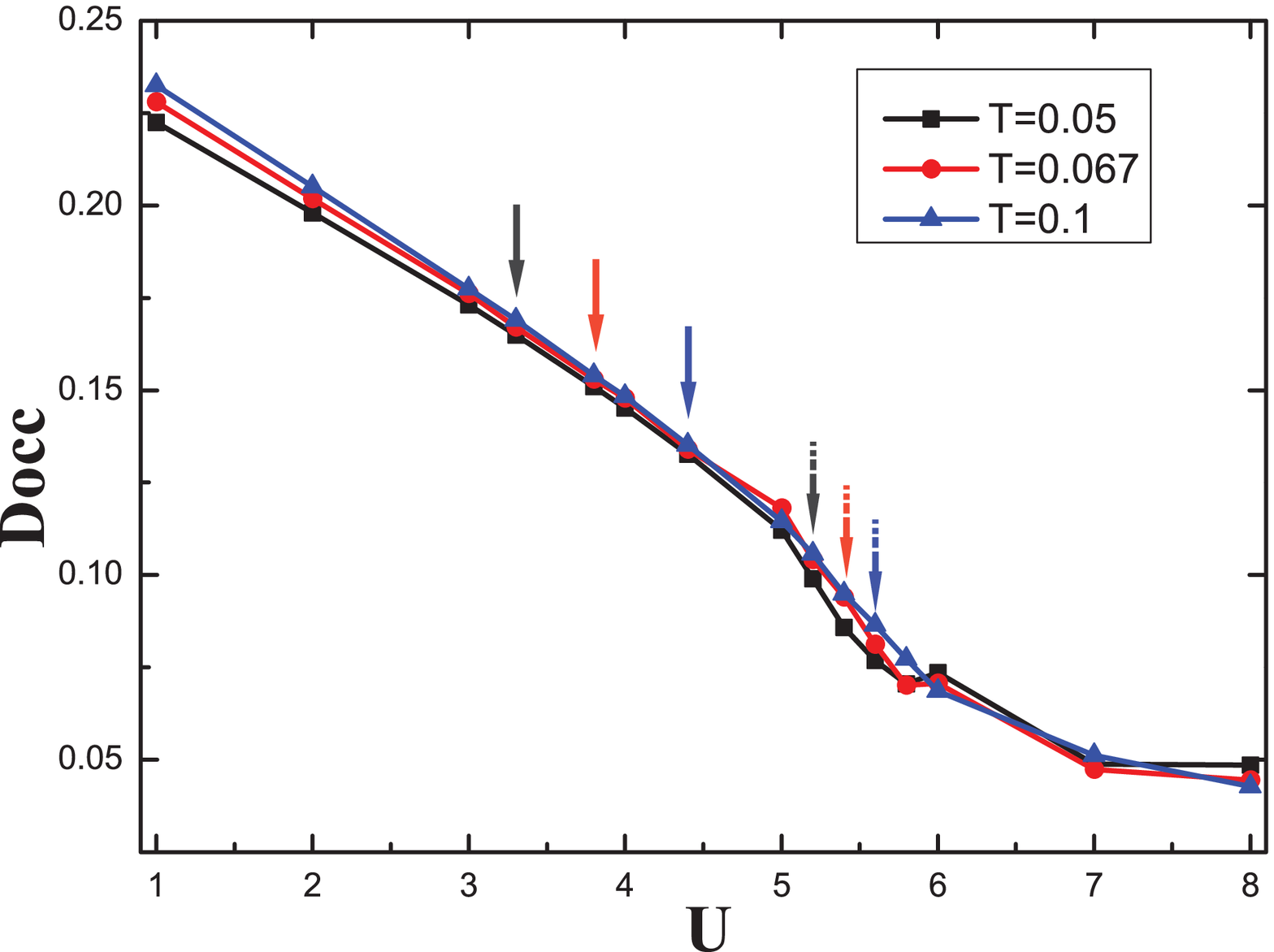,width=11cm}
    \end{center}
    \label{fig:CN}
\end{figure}

\begin{figure}
    \begin{center}
        \epsfig{file=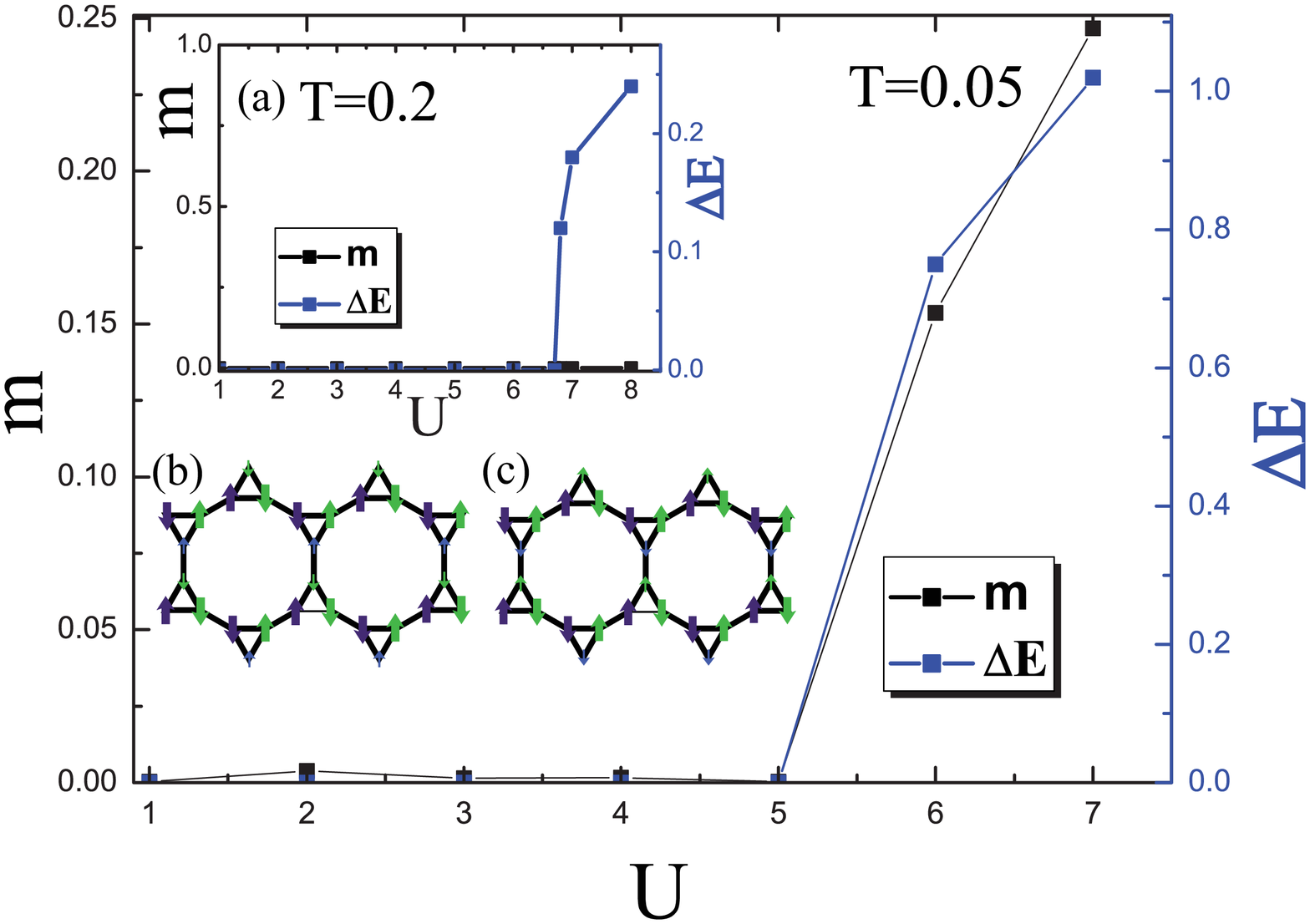,width=12cm}
    \end{center}
    \label{fig:Vband}
\end{figure}
\newpage
\begin{figure}
    \begin{center}
        \epsfig{file=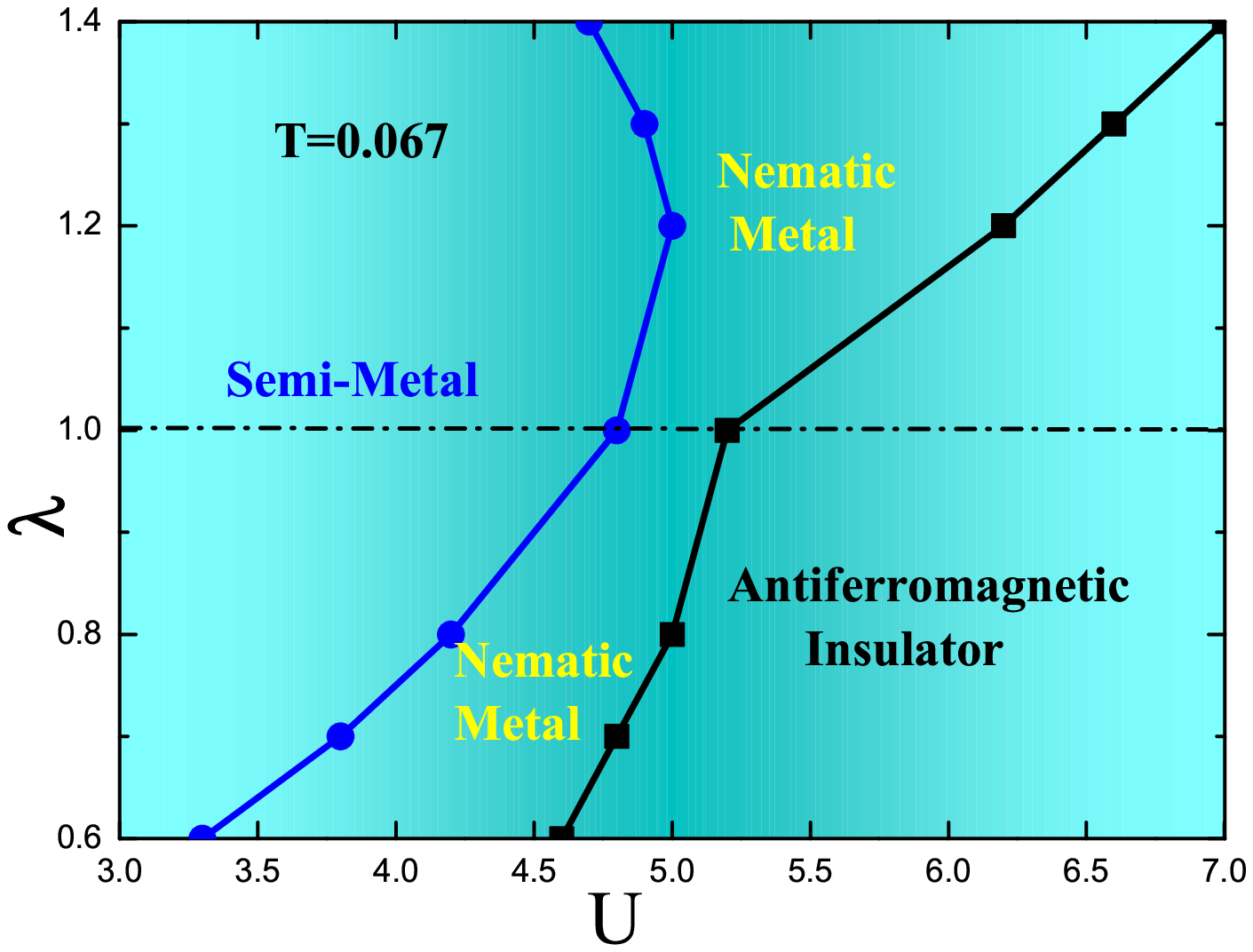,width=15cm}
    \end{center}
    \label{fig:BC}
\end{figure}

\end{document}